\documentclass[prl,twocolumn,showpacs,amsmath,preprintnumbers,amssymb,superscriptaddress,letterpaper]{revtex4}

\usepackage{epsfig}
\usepackage{amsmath}
\usepackage{amssymb}
\usepackage{amsfonts}
\usepackage{graphicx}

\begin{document}
\title{Efficiency of discrete-time ratchets}

\author{Pau Amengual}
\email{pau@imedea.uib.es}
\affiliation{Instituto Mediterr\'aneo de Estudios Avanzados, IMEDEA (CSIC-UIB),\\
Ed.~Mateu Orfila, Campus UIB, E-07122 Palma de Mallorca, Spain}
\affiliation{Centre for Biomedical Engineering (CBME)
and \\ Department of Electrical and Electronic Engineering,\\
The University of Adelaide, SA 5005, Australia}
\author{Ra\'ul Toral}
\affiliation{Instituto Mediterr\'aneo de Estudios Avanzados, IMEDEA (CSIC-UIB),\\
Ed.~Mateu Orfila, Campus UIB, E-07122 Palma de Mallorca, Spain}
\author{Andrew Allison}
\affiliation{Centre for Biomedical Engineering (CBME)
and \\ Department of Electrical and Electronic Engineering,\\
The University of Adelaide, SA 5005, Australia}
\author{Derek Abbott}
\affiliation{Centre for Biomedical Engineering (CBME)
and \\ Department of Electrical and Electronic Engineering,\\
The University of Adelaide, SA 5005, Australia}

\begin{abstract}
Recently there has been much interest in discrete forms of Brownian ratchets, 
using a game-theoretic formalism. Using the approach pioneered by Parrondo, we develop 
a new method for obtaining the stationary probabilities and probability current for the 
case of discrete-time and discrete-space ratchets. We then use this result to calculate 
the Parrondian ratchet efficiency in two cases: firstly, for Parrondo's original 
system and, secondly, for a set of probabilities derived from a discretized ratchet potential. 
\end{abstract}


\maketitle

Since the field of  Brownian ratchets acquired its importance, there have been numerous studies on the energetics of these microscopic devices~\cite{energetics,KHT98}. Parrondo~\cite{games} devised a discrete-time and discrete-space version of the flashing ratchet model~\cite{flashingratchet} within a game-theoretic framework, and recent work has established the connection between game probabilities 
and physical variables that describe the motion of a Brownian particle~\cite{discrete}. For extensive reviews of this area that show the relationship of the Parrondo effect to classical random walks, quantum walks, fractal pattern formation and spin systems, for example, see~\cite{games}.
The significance of game-theoretic approaches in physical systems has recently been reviewed~\cite{abbott02} and the Parrondo effect in transport processes has recently been discussed~\cite{heath02}. There is also emerging interest in the so-called continuous-discrete interface~\cite{argwala02}, and the Parrondo approach is also of relevance in that arena.

The original Parrondo's paradox involves the alternation of two games. The first game, game A, 
has a winning probability given by $p=\frac{1}{2}-\epsilon$. It is easy to check that for $\epsilon=0$ 
game A is fair, whereas for small $\epsilon$ game A turns into a losing game.  Game B is a capital 
dependent game, where the winning probability depends on the capital of the player being multiple 
of three or not. If the capital is multiple of three the winning probability is $p_1=\frac{1}{10}-\epsilon$, otherwise it is $p_2=\frac{3}{4}-\epsilon$. As in the case of game A, for $\epsilon=0$ game B is 
a fair game, and a losing one for $\epsilon>0$. This set of probabilities can be summarized as, 
\vspace{-0.1truecm}
\begin{equation}\label{probabilities_parrond}
\vspace{-0.1truecm}
    p=\frac{1}{2}-\epsilon,~~~ p_1=\frac{1}{10}-\epsilon,~~~ p_2=\frac{3}{4}-\epsilon.
\end{equation} 
Parrondo conjectured that this system is a discrete form of the flashing ratchet, and this has 
recently been rigorously demonstrated within a Fokker-Planck framework~\cite{AA02}. 
However, finding the correct formalism for describing the efficiency of the discrete ratchet 
and relating it back to the continuous case, has been problematic~\cite{harmer01}---this has 
motivated a solution, which is now presented for the first time in this paper.

The flashing ratchet consists of a Brownian particle under the influence of a potential that can be switched on and off either stochastically or periodically. Its dynamics can be described through the following Langevin equation, $\dot{x}=-V'[x(t)]\cdot\zeta(t)+F_{\mathrm{ext}}+ D[x(t)]\cdot\xi(t)$, where (i) $\xi(t)$ accounts 
for white noise, (ii) $\zeta(t)$ is a form of dichotomous noise that switches on (state B) and off (state A) the potential $V(x)$, and (iii) $F_\mathrm{ext}$ is a force acting on the particle, that can be exerted by an external agent. It is required that the potential has a certain degree of spatial asymmetry in order to 
obtain directed motion from these fluctuations. Usually a ratchet-like potential is used, 
\vspace{-0.1truecm}
\begin{equation}\label{ratchetpotential}
\vspace{-0.1truecm}
V(x)= V_0 \left({\sin }\left(\frac{2\pi x}{L}\right)+ \frac{1}{4}{\sin}\left(\frac{4\pi x}{L}\right)\right),
\end{equation} 

 where $V_0$ denotes the amplitude of the potential and $L$ its spatial periodicity. Although other similar potentials perform the same task, Eq.~(\ref{ratchetpotential}) is convenient for analytical purposes. 

The corresponding set of Fokker-Planck equations related to the Langevin equation
that describes the transitions of the particle between states A and B are~\cite{flashingratchet},
\vspace{-0.2truecm}
\begin{eqnarray*} 
\vspace{-0.1truecm}
\textstyle{\frac{\partial P_A(x,t)}{\partial x}=-\frac{\partial(\mathcal{J}_A(x,t)
P_A(x,t))}{\partial x}}&-&\omega_{A\rightarrow B}P_A(x,t)\nonumber\\
&+&\omega_{B\rightarrow A} P_B(x,t) \\
\textstyle{\frac{\partial P_B(x,t)}{\partial x}=-\frac{\partial(\mathcal{J}_B(x,t)
P_B(x,t))}{\partial x}}&-&\omega_{B\rightarrow A}P_B(x,t)\nonumber\\
&+&\omega_{A\rightarrow
B}P_A(x,t), 
\end{eqnarray*}

where $P_A(x,t)$ (resp.~$P_B(x,t)$) denotes the probability of finding the particle in state A (resp.~B) at a given position $x$ and time $t$. The term
$\omega_{\alpha\rightarrow \beta}$ accounts for the transition rate between state $\alpha$ and $\beta$. The probability currents $\mathcal{J}_A$ and $\mathcal{J}_B$
are given by
\vspace{-0.2truecm}
\begin{eqnarray*}
\vspace{-0.1truecm}
\mathcal{J}_A(x,t) &=& F_{\mathrm{ext}}P_A(x,t)-\textstyle{\frac{\partial (D(x)P_A(x,t))}
{\partial x}}\\
\mathcal{J}_B(x,t) &=& \textstyle{\left(F_{\mathrm{ext}}-\frac{\partial V(x)}{\partial x}\right)
P_B(x,t)-\frac{\partial (D(x)P_B(x,t))}{\partial x}}. 
\end{eqnarray*}

In the stationary regime we have $P_A(x,t)=P_A(x)$ and $P_B(x,t)=P_B(x)$. The current in this regime is constant~\cite{flashingratchet} and given by $\mathcal{J}=\mathcal{J}_A(x)+\mathcal{J}_B(x)$. With this in mind, and using the same notation as in~\cite{discrete}, we propose a model that describes the dynamics of the capital of the player when alternating between games A and B. The set of Master Equations (ME) 
describing this process are
\vspace{-0.1truecm}
\begin{eqnarray}
\vspace{-0.1truecm}
P^A_i(\tau+1)&=&(1-\omega_{AB})[a_{-1}^i P^A_{i-1}(\tau)+a_{0}^i
P^A_{i}(\tau)\nonumber\\
+a_{1}^iP^A_{i+1}(\tau)]&+&\omega_{BA}[b_{-1}^i
P^B_{i-1}(\tau)+b_{0}^iP^B_{i}(\tau)\nonumber\\
+b_{1}^iP^B_{i+1}(\tau)]&&\label{4a}\\
P^B_i(\tau+1)&=&(1-\omega_{BA})[b_{-1}^iP^B_{i-1}(\tau)+b_{0}^i
P^B_{i}(\tau)\nonumber\\
+b_{1}^iP^B_{i+1}(\tau)]&+&\omega_{AB}[a_{-1}^i
P^A_{i-1}(\tau)+a_{0}^iP^A_{i}(\tau)\nonumber\\
+a_{1}^iP^A_{i+1}(\tau)].&&\label{4b}
\end{eqnarray}

\begin{figure}[t]
\centerline{\epsfig{figure=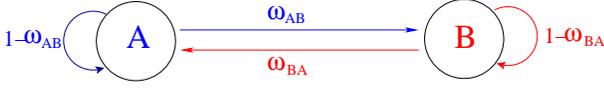,width=8cm}} 
\caption{\label{statescheme} State transition diagram between games A and B.}
\vspace{-0.22truecm}
\end{figure}

The terms $P^A_{i}(\tau)$, $P^B_{i}(\tau)$ account for the probability that the player plays game A or B with a capital $i$ at time $\tau$ respectively. Also the term 
$a_{-1}^i$ (resp.~$b_{-1}^i$) denotes the probability of winning if the player plays game A (resp.~B) with a capital $(i-1)$; $a_{0}^i$ (resp.~$b_{0}^i$) is the
so--called self--transition probability, that is, the probability that the player will remain with the same capital after a round played for a given capital $i$; and 
finally $a_{1}^i$ (resp.~$b_{1}^i$) is the probability of losing when the player has a capital $(i+1)$ and is playing game A (resp.~B). The transitions between
game states are shown in Fig.~\ref{statescheme}.

In order to preserve normalization, Eqs.~(\ref{4a}) \&~(\ref{4b}) must accomplish the condition $\sum_i[P^A_{i}(\tau)+P^B_{i}(\tau)]=1$ for the stationary 
probabilities, and $a_{-1}^{i+1}+a_0^i+a_1^{i-1}=1$ together with $b_{-1}^{i+1}+b_0^i+b_1^{i-1}=1$ for the transition probabilities in games A and B respectively. 
Making use of the normalization condition for the transition probabilities, Eqs.~(\ref{4a}) \&~(\ref{4b}) can be rewritten in the form of continuity equations for 
$P_A^i(\tau)$ and $P_B^i(\tau)$ as
\vspace{-0.1truecm}
\begin{eqnarray*}
\vspace{-0.1truecm}
P^A_i(\tau+1)-P^A_i(\tau)&=&(1-\omega_{AB})[a_{-1}^iP^A_{i-1}(\tau)-\nonumber\\
(a_{1}^{i-1}+a_{-1}^{i+1}) P^A_{i}(\tau)&+&a_{1}^iP^A_{i+1}(\tau)]-\omega_{AB}P_i^A(\tau)+\nonumber\\
+\omega_{BA}[b_{-1}^iP^B_{i-1}(\tau)&+&b_{0}^iP^B_{i}(\tau)+b_{1}^iP^B_{i+1}(\tau)]\\
P^B_i(\tau+1)-P^B_i(\tau)&=&(1-\omega_{BA})[b_{-1}^iP^B_{i-1}
(\tau)-\nonumber\\
(b_{-1}^{i+1}+b_{1}^{i-1}) P^B_{i}(\tau)&+&b_{1}^iP^B_{i+1}(\tau)]-\omega_{BA}P^B_i(\tau)+\nonumber\\
+\omega_{AB}[a_{-1}^iP^A_{i-1}(\tau)&+&a_{0}^iP^A_{i}(\tau)+a_{1}^iP^A_{i+1}(\tau)] 
\end{eqnarray*}

where the \emph{lhs} of both expressions must equal zero, in the stationary regime.  These equations, combined with the normalization condition for the stationary probabilities, form a set of equations that must be solved for the variables $P_i^A$, $P_i^B$ for each $i$. For the case of Parrondo's games it is necessary only to consider the $0$, $1$ and $2$ states due to the presence of a modulo three periodicity. Therefore, in matrix form we must solve the $\mathbb{C}\cdot\mathbb{P}=\mathbb{U}$, where $\mathbb{P}=\{P_0^A,P_1^A,P_2^A,P_0^B,P_1^B,P_2^B\}^T$, $\mathbb{U}=\{1,1,1,1,1,1\}^T$, and

\vspace{-0.1truecm}
\begin{widetext}
\begin{displaymath}
\vspace{-0.1truecm}
\mathbb{C}=\left(
{\begin{array}{cccccc}
1-A_0-\omega_{AB}&1+(1-\omega_{AB})q_1^A &1+(1-\omega_{AB})p_2^A 
&1+\omega_{BA} r_0^B &1+\omega_{BA}q_1^B &1+\omega_{BA}p_2^B\\

1+(1-\omega_{AB})p_0^A &1-A_1-\omega_{AB} &1+(1-\omega_{AB})q_2^A 
&1+\omega_{BA}p_0^B &1+\omega_{BA}r_1^B &1+\omega_{BA}q_2^B\\

1+(1-\omega_{AB})q_0^A &1+(1-\omega_{AB})p_1^A &1-A_2-\omega_{AB} 
&1+\omega_{BA}q_0^B &1+\omega_{BA}p_1^B &1+\omega_{BA}r_2^B \\

1+\omega_{AB}r_0^A &1+\omega_{AB}q_1^A &1+\omega_{AB}p_2^A 
&1-B_0-\omega_{BA} &1+(1-\omega_{BA})q_1^B &1+(1-\omega_{BA})p_2^B \\

1+\omega_{AB}p_0^A &1+\omega_{AB}r_1^A &1+\omega_{AB}q_2^A 
&1+(1-\omega_{BA})p_0^B &1-B_1-\omega_{BA} &1+(1-\omega_{BA})q_2^B \\

1+\omega_{AB}q_0^A &1+\omega_{AB}p_1^A &1+\omega_{AB}r_2^A 
&1+(1-\omega_{BA})q_0^B &1+(1-\omega_{BA})p_1^B &1-B_2-\omega_{BA}\\
\end{array}}
\right),
\end{displaymath}
\end{widetext}
and where $A_i=(1-\omega_{AB})(p_i^A+q_i^A)$ and $B_i=(1-\omega_{BA})(p_i^B+q_i^B)$. The solution can be obtained through $\mathbb{P}=\mathbb{C}^{-1}\cdot\mathbb{U}$.
The analytical solutions for $P_i^A$ and $P_i^B$ are too lengthy to be presented here---however, we will show some results concerning the original Parrondo's games A 
and B. As we already stated before, these games are alternated using a \emph{mixing} probability $\gamma$. This means that independently of the previously played 
game, we have always a probability $\gamma$ of playing game A  and a probability $1-\gamma$ of playing game B, in the next round. It can be easily checked that the 
latter condition is equivalent to $\omega_{AB}=1-\gamma$ and $\omega_{BA}=\gamma$.

Making use of the previous expressions for $\omega_{AB}$, $\omega_{BA}$ together with the set of probabilities already defined in Eq.~(\ref{probabilities_parrond}), 
and by letting $\eta=-169 + 11\,{\gamma }^2 + 16\,\epsilon  - 240\,{\epsilon }^2 - 2\,\gamma \,\left( 11 + 8\,\epsilon  \right)$, we obtain the following 
expressions for the stationary probabilities for game A and B
\vspace{-0.1truecm}
\begin{align}
&P^A_0=\textstyle{\frac{-5\,\gamma \,\eta_2}{\eta}}&
&P^A_1=\textstyle{\frac{2\,\gamma \,\eta_3}{\eta}}&
&P^A_2=\textstyle{\frac{2\,\gamma \,\eta_4}{\eta}}&\\
&P^B_0=\textstyle{\frac{5\,\left(\gamma-1 \right) \,\eta_2}{\eta}}&
&P^B_1=\textstyle{\frac{-2\,\left(\gamma-1 \right) \,\eta_3}{\eta}}&
&P^B_2=\textstyle{\frac{-2\,\eta_4}{\eta}}.&
\end{align}

where $\eta_2=\left(13+{\gamma }^2-8\epsilon+16{\epsilon}^2+2\gamma\left(-1+4\epsilon\right)\right)$, $\eta_3=\left(-13+4{\gamma}^2-6\epsilon-40{\epsilon}^2+3\gamma\left(-7+2\epsilon \right)\right)$ and $\eta_4=\left(-39+4{\gamma}^2-6\epsilon-40{\epsilon}^2 +\gamma\left(5+6\epsilon \right)  \right)$

\begin{figure}[t] 
\centerline{\epsfig{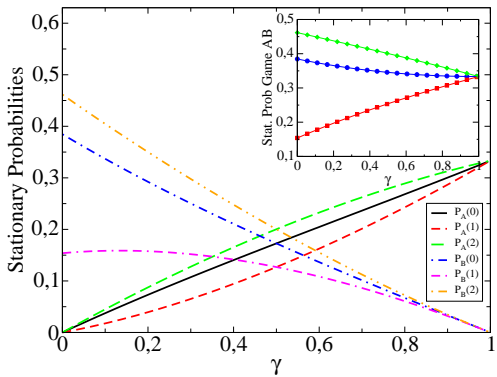}} 
\caption{\label{Probabilities}Plot of the evolution of the stationary probabilities for game A and game B versus the mixing variable $\gamma$. The inset shows that the sum of both probabilities, $P^T_i=P_i^A+P_i^B$, agrees with the expressions obtained for the stationary probabilities $\Pi_0$, $\Pi_1$, $\Pi_2$ obtained for the mixed game AB.}
\vspace{-0.20truecm}
\end{figure}

They are expressed in terms of the mixing probability $\gamma$ and the biasing term $\epsilon$. In Fig.~\ref{Probabilities} we plot both stationary probabilities $P^A_i$, $P^B_i$ versus the mixing probability $\gamma$ for the simplest case $\epsilon=0$. The inset in Fig.~\ref{Probabilities} shows the total probability $P^T_i=P^A_i+P^B_i$ for each state, that is, the probability of finding the capital of the player modulo three $(P^T_0,P^T_1,P^T_2)$ considering both games A and B. These values are compared with the stationary probabilites $\Pi_0$, $\Pi_1$, and $\Pi_2$ obtained through a Discrete-Time Markov Chain (DTMC) analysis~\cite{LAA03}. Fig.~\ref{Probabilities} confirms that both probability sets fully coincide. 

The total current, as in the continuous case, will be constant in value (that is,
independent of capital $i$), $J=J_i^A+J_i^B$, where $J_i^A=p_i^A\cdot P_i^A-q_{i+1}^A\cdot P_{i+1}^A$ and
$J_i^B=p_i^B\cdot P_i^B-q_{i+1}^B\cdot P_{i+1}^B$. For obtaining the average gain, we must multiply the current $J$ with the periodicity $M$ of the games, which in our case is $M=3$, and so $G=3J$. Simulations reveal excellent agreement with a theoretical
plot of $G$ versus $\gamma$.

Now we have a method for obtaining the stationary probabilities $P_i^A$, $P_i^B$, we turn to the problem of evaluating efficiency for our discrete-time system. But firstly, we introduce a result from~\cite{discrete}, where a method is presented for obtaining a potential, given a set of probabilities defining a game. Its most 
important property is that the potential obtained is unbiased if a game is fair, and biased otherwise (with a positive slope if the game is losing and negative if winning). The equation used for calculating the potential is,
\vspace{-0.1truecm}
\begin{equation}\label{potential}
\vspace{-0.1truecm}
V_i=-\sum_{j=1}^{i}\ln\left(\frac{\frac{p_{j-1}}{1-r_{j-1}}}
{\frac{1-p_j-r_j}{1-r_j}}\right).
\end{equation}

It can be checked that if our probability set $\{p_1,p_2,\ldots,p_{L-1}\}$ describes a fair game, that is, $\prod_{i=1}^{L-1}p_i =\prod_{i=1}^{L-1}q_i$, then the potential is periodic $V_0 = V_L$. It also works the other way around. If we have a discretized potential $V_i$, and we are interested in obtaining its related 
probabilities, we can make use of  Eq.~(\ref{potential}) solving for the probabilities $\{p_i,r_i,q_i\}$ (see~\cite{discrete} for further details). 

This previous result allows us to evaluate the efficiency in two different cases. On one hand, given two probability sets defining games A and B, we will obtain their related potentials using Eq.~(\ref{potential}) and then the efficiency can be derived. On the other hand, we can calculate the efficiency given two potentials, from which we can obtain their respective probability sets inverting Eq.~(\ref{potential}). For the latter case, we use a flat potential for state A, that is, $V_A=0$, and state B, $V_B$ will be obtained from Eq.~(\ref{ratchetpotential}) with $L=3$.

In~\cite{DBA99} a new expression for the efficiency is presented. It is based on a new definition of the energy output $\mathcal{E}_{\mathrm out}$: `\ldots We define the energy output $\mathcal{E}_{\mathrm out}$ of an engine as the \emph{minimum} energy input $\mathcal{E}_{\mathrm in}$ required to accomplish the same task as the engine.' The novelty of this definition is that permits the evaluation of the efficiency for a Brownian particle even in the absence of an external load $F$. So we can evaluate the energy output (or   $\mathcal{P}_{\mathrm out}$) using the  expression, $\mathcal{P}_{\mathrm{out}}=\mathcal{P}_{\mathrm{in}}^{\mathrm{min}}=F_{\mathrm{ext}}\cdot v +\gamma v^2$.
Recalling that in our system, $v=J\cdot L$ and $\gamma=1$,  we obtain $\mathcal{P}_{\mathrm{out}}=F_{\mathrm{ext}} J L + J^2 L^2$. This is the equation to be used for determining the energy output of our system.

The energy input of the system is the energy that we must supply to the system when switching between the two potentials. For evaluating this energy input in our system we need a potential related to each of the two games. Therefore, if we are dealing with probabilities defining our games A and B, we will make use of 
Eq.~(\ref{potential}) for obtaining the potential for each game. The energy input can be calculated theoretically by means of a probability flux balance. In the stationary regime, the net flux from a given game, say game A, and state $i$, towards the other game B and the same state $i$ can be calculated through the difference equation $J^{\mathrm{net}}_{AB}(i)= J_{i-1}^A-J_{i}^A=(p_{i-1}^AP_{i-1}^A-q_{i}^AP_{i}^A)-(p_i^AP_i^A-q_{i+1}^A P_{i+1}^A).$
Clearly the net current $J^{\mathrm{net}}_{AB}(i)$ equals the opposite current from game B to game A, that is, $\vert J^{\mathrm{net}}_{AB}(i)\vert=\vert J^{\mathrm{net}}_{BA}(i)\vert$, where $ J^{\mathrm{net}}_{BA}(i)= J_{i}^B-J_{i-1}^B$. The input power can now be obtained and is given by $\mathcal{P}_{\mathrm{input}}=\sum_i^{L-1}J^{\mathrm{net}}_{AB}(i)\cdot(V_B(i)-V_A(i))$.
For the simplest case when $F_\mathrm{ext}=0$, for the original Parrondo games, we obtain $\eta= 1.1078\cdot 10^{-2}$ for $\gamma=0.36$. For the alternation with the potentials $V_A$ and $V_B$, we obtain a maximum value for the efficiency of $\eta= 1.061\cdot 10^{-3}$ when $\gamma=0.35 $. The system possesses a low efficiency mainly because it works in an irreversible manner, far from its equilibrium state. The magnitude obtained for the efficiency agrees with other studies for the \emph{on-off} ratchets \cite{energetics,onoff}. Now that we have obtained an efficiency metric, it now provides a basis for comparing different Parrondian or discrete-time ratchets, and provides a basis in the search for higher efficiency systems.  

Now we evaluate the efficiency, when $F_{\mathrm{ext}}\neq 0$, which is difficult when considering alternation between games A and B. Clearly, the applied external force $F_{\mathrm{ext}}$ must bias the potential with a slope equal to $-F_{\mathrm{ext}}$. Whereas it can be demonstrated that the biasing  parameter, $\epsilon$, 
introduced earlier in Eq.~(\ref{probabilities_parrond}), is exactly equivalent to $F_{\mathrm{ext}}$ only for the case of game A---due to the non-constant values of the probabilities in case of game B, $\epsilon$ is no longer equivalent to the force. 

\begin{figure}[t]
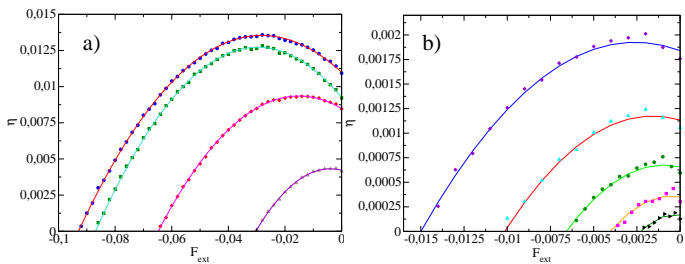

\makebox{\centerline{\epsfig{figure=Fig3a.eps,width=4.5cm}\epsfig{figure=Fig3b.eps,width=4.5cm}}} \caption{\label{eff_parr}Plots of the efficiency versus the external forcing $F_\mathrm{ext}$ for two cases: a) Alternating the original Parrondo's games A and B with different mixing probability $\gamma$. From top to bottom: $\gamma=0.4,0.2,0.6,0.8$. b) Alternating between the potentials $V_A=0$ and $V_B$ for a value of $\gamma=0.45$. The plot includes the curves for different values of the amplitude $V_0$ for the potential $V_B$, from top to bottom: $V_0=0.45,0.4,0.35,0.3,0.25$. For both graphs the theoretical curves are the solid lines and the numerical values are the circles.}
\vspace{-0.20truecm}\end{figure}

Because of this, we proceed as follows: given a set of probabilities $\{p_i,r_i,q_i\}$ defining our game B, through Eq.~(\ref{potential}) we calculate its discretized potential $V_i$. Then, for $F_\mathrm{ext}\neq 0$ we add this \emph{extra} bias through $V^{\mathrm{new}}_i=V_i-F_{\mathrm{ext}}\cdot i $, where $i$ 
denotes the state. Finally the new set of probabilities will be obtained inverting Eq.~(\ref{potential}), using the new values for the potential $\{p^{\mathrm{n}}_i,r^{\mathrm{n}}_i,q^{\mathrm{n}}_i\}$. If we want to calculate the efficiency when alternating between the potentials $V_A$ and $V_B$, we need only to add the slope to these potentials and then invert Eq.~(\ref{potential}) for obtaining the probability sets.
In Fig.~\ref{eff_parr}a, the efficiency for the original Parrondo's games is evaluated for different values of the mixing probability $\gamma$. It can be appreciated that the efficiency attains a maximum value for a value of $F_{\mathrm{ext}}\neq 0$, that corresponds to a lower value for the current than in the case of $F_{\mathrm{ext}}= 0$. This effect also has been found in other models, for example in~\cite{onoff}. In Fig.~\ref{eff_parr}b we compare theoretical and numerical simulations for the efficiency when alternating between the potentials $V_A$ and $V_B$. We have obtained the efficiency for different values of the 
potential amplitude $V_0$ for the potential $V_B$. As in the previous case, the efficiency also attains a maximum for a value of $F_{\mathrm{ext}}\neq 0$.

In conclusion, for the first time we provide a method that permits an analytical solution for the stationary probabilities $P_i^A$, $P_i^B$, and the average gain $G=3J$ of a player that alternates between two games A and B with different transition probabilities $\omega_{AB}$, $\omega_{BA}$. Using this approach, we then evaluated the efficiency in the case of the original Parrondo games, as well as for a given set of potentials $V_A$, $V_B$, both of them sharing the essential features with the continuous model of a flashing Brownian ratchet. 

This work was supported by GTECH Australasia; the Ministerio de Ciencia y Tecnolog\'{i}a (Spain) and FEDER, projects BFM2001-0341-C02-01 and BFM2000-1108; P.A. 
acknowledges support from the Govern Balear, Spain.

\end{document}